\documentclass{article}

\usepackage{PRIMEarxiv}

\usepackage[utf8]{inputenc} % allow utf-8 input
\usepackage[T1]{fontenc}    % use 8-bit T1 fonts
\usepackage{hyperref}       % hyperlinks
\usepackage{url}            % simple URL typesetting
\usepackage{booktabs}       % professional-quality tables
\usepackage{amsfonts}       % blackboard math symbols
\usepackage{nicefrac}       % compact symbols for 1/2, etc.
\usepackage{microtype}      % microtypography
\usepackage{lipsum}
\usepackage{fancyhdr}       % header
\usepackage{graphicx}       % graphics
\graphicspath{{media/}}     % organize your images and other figures under media/ folder
\usepackage{gensymb}
\usepackage{textcomp}
\usepackage{comment}

\title{Beamforming using Digital Piezoelectric MEMS Microphone Array
}

\author{
  Ricky Leman, Ben Travaglione, and Melinda Hodkiewicz* \\
  School of Engineering, \\
  The University of Western Australia, \\
  Perth, Australia\\
  \texttt{*melinda.hodkiewicz@uwa.edu.au} \\
  }

\begin{document}
\maketitle

\begin{abstract}
The recent explosion in low-cost, low-power wireless microcontrollers, coupled with low-power, robust MEMS sensors has opened up the opportunity to create new forms of low-cost Industrial Internet-of-Things (IIoT) devices for condition monitoring.
Piezoelectric MEMS microphones constructed with a cantilever diaphragm are a potential solution against failure modes, such as water and dust ingress, that have challenged the use of capacitive MEMS microphones in industrial applications. 
In this paper, we couple a pair of piezoelectric MEMS microphones to a COTS microcontroller to create a stand-alone microphone array capable of discerning the direction of a noise source. 
The microphone array is designed to acquire sound data without aliasing at frequencies of 2000 Hz or below. Testing is conducted in an anechoic chamber.
We compare the performance of this microphone array to a simple idealized theoretical model. 
The experimental results obtained in the anechoic chamber compare well with the theoretical model.
The work stands as a proof-of-principle.
By providing detailed information on how we coupled the sensors to a COTS microcontroller, and the open-source code used to process the data, we hope that others will be able to build upon this work by expanding on both the number and type of sensors used.
\end{abstract}

% Keywords

\section{Introduction}
\label{sec:introduction}
%{C}{ondition} monitoring involves analysing equipment parameters in operation to track the health of machinery and inform predictive maintenance. It is fundamental to condition-based maintenance strategies in industry due to its ability to detect potential failures. Several methods of condition monitoring exist, of which vibration monitoring is widely used within the resources industry \cite{tandon1999review}. The method of acquiring vibration data requires sensors to be manually placed onto a system. This involves exposing personnel to moving machinery, high temperatures, and hazardous electrical contact.  

Acoustic analysis offers the ability to perform condition monitoring without the need for surface contact on a machine. Sound emissions from machinery are acquired by microphone arrays and processed using a signal processing technique known as beamforming. This enables detection of the source location of a desired signal amidst noise. This method of acoustic analysis has been demonstrated in literature through underwater hydrophone applications \cite{fox2001monitoring} to medical applications such as hearing aids \cite{welker1997microphone}, and in the resources industry for leak detection \cite{anastasopoulos2009acoustic} and fan bearing issues \cite{oh2011estimation}. 

Use-cases of acoustic analysis conventionally utilise capacitive microphones in their design. This is due to their high sensitivity, flat frequency response and low noise level \cite{shah2019design}. However, sensors used within heavy industry are subject to high levels of dust exposure and water ingress. These are environmental factors that have a known impact on the performance of capacitive MEMS microphones \cite{samper2003mems,Chew2017mems}, rendering them potentially unsuitable for chemical, mining or oil and gas applications. Piezoelectric MEMS microphones provide a potential solution to the harsh environmental conditions on-site within the resources industry. Their cantilever diaphragm removes the need for a back-plate found in capacitive microphones, resulting in a waterproof and dustproof design \cite{Vesper2020VM3000}. The VM3000 produced by Vesper Technologies is a commercially-available digital piezoelectric MEMS microphone that has the potential for industrial application under harsh environments. However, there is a lack of published literature regarding the beamforming performance of the VM3000s MEMS microphones. The aim of this paper is to describe the design and assess the beamforming performance of a stand-alone digital microphone array with Wi-Fi connectivity using the VM3000 piezoelectric MEMS microphones.   

%A test environment featuring a signal generator, speaker, and microphone array provides a controllable, known output sound signal. Python is used to control the signal generator, microphone array and to process experimental data. The microphone array prototype is used to record a linear frequency sweep test signal of 500 to 3000 Hz between array angles of 0\degree {} to 360 \degree {} at increments of 20 \degree. Test data is processed using time-domain delay-sum beamforming to generate real beam patterns. 
   
The paper is organised as follows. Section 2 explores past research on condition monitoring using sound analysis, fundamental theory underpinning microphone arrays,  and a comparison between piezoelectric and capacitive MEMS microphones. Section 3 outlines the engineering design process of the hardware and software aspects of this work. Section 4 describes the build process of the microphone array and Section 5 describes the method of testing to assess beamforming performance. Section 6 discusses beamforming performance results from testing, Section 7 concludes the paper with a recap of key takeaways and outlines recommendations for future work.

\section{Background}
\label{sec:literature review}
\subsection{Condition Monitoring with Acoustic Analysis}
%In 2019 the global Condition Monitoring market received a valuation of USD 3.09 billion and is expected to reach a value of USD 5.15 billion by 2026 at a compounded annual growth rate (CAGR) of 9.02\% \cite{MachineConditionMonitoringMarket1,MachineConditionMonitoringMarket2}. The ability to conduct predictive maintenance on equipment as informed by condition monitoring reduces maintenance and replacement cost whilst improving equipment safety and reliability.  It is a widely adopted technique as demonstrated through the existence of several condition monitoring standards such as ISO 13373 and ISO 13381 \cite{hitchcock2006iso}. Although vibration analysis is predicted to capture the majority of market share as seen by its extensive range of applications \cite{randall2011vibration}, installation of vibration sensors incur high implementation costs and are intrusive in nature \cite{zhou2007bearing,albarbar2017mems}.

Sound analysis for condition monitoring involves the acquisition of sound data using microphones and subsequent signal conditioning where sound is amplified and digitised. Sound data can then be converted into the frequency domain through  Fourier Transform methods for comparison against known sound signatures. Several use-cases of this method have been demonstrated in the monitoring of mechanical elements such as internal-combustion engine wear \cite{delvecchio2018vibro}, induction motor bearing faults \cite{garcia2012application}, through to railway obstruction fault detection applications \cite{lee2016fault}. These use cases demonstrated high fault classification accuracy using standard vibration analysis methods as a benchmark for comparison. 

\subsection{Microphone Arrays and Beamforming}
Microphone array is a term used to denote a sound acquisition device that is comprised of multiple microphones arranged in various geometrical layouts for beamforming. Beamforming is a signal processing technique that enables discrimination of different signals based on the physical location of sound sources relative to a microphone array \cite{brandstein2013microphone}. The combination and processing of various microphone outputs allow received signals to be cleaned from contaminating interference and ambient noise \cite{dmochowski2008spatial}. As a technology which enables sound source localisation, it has seen commercial applications in videoconferencing technology \cite{arcondoulis2010design}, communications infrastructure \cite{ioannides2005uniform} and hearing aids \cite{welker1997microphone,brandstein2013microphone}. The application of microphone arrays and beamforming as a method for condition monitoring is not widely adopted in industry. However, several papers evaluating its use-cases are available in literature \cite{cabada2017fault,orman2013usage}.     

The geometrical design of a microphone array directly influences its achievable beamforming resolution \cite{prime2013comparison}. A long-standing issue with regularly spaced microphones is spatial aliasing, where spatial sampling must occur at a rate greater than half the wavelength of incoming sinusoids to produce an unambiguous beam pattern \cite{dmochowski2008spatial}. This is achieved through an inter-microphone spacing compliant with:

\begin{equation}
    \label{Inter-microphone distance restriction}
    d < \frac{\lambda_\mathrm{min}}{2} \; \mathrm{where} \; \lambda_\mathrm{min} = \frac{c}{f_\mathrm{max}} 
\end{equation}

Where $\lambda_\mathrm{min}$ is the minimum wavelength of the signal of interest, $c$ is the speed of sound (around 343 ms$^{-1}$ in air) and $f_\mathrm{max}$ is the maximum frequency of the signal of interest \cite{brandstein2013microphone}.

Spatial aliasing results in an inability to distinguish the direction of arrival of a desired signal, producing ghost sources in beam patterns of comparable amplitude to the true signal source direction \cite{chiariotti2019acoustic}. As such, many irregular array designs with unique inter-microphone spacing known as non-redundant array designs have been explored in literature \cite{prime2013comparison}. These designs, however, are complex in comparison to the traditional uniform linear array (ULA) design, the applications of which have been documented in several publications \cite{ottoy2016low,thoen2017ultra,chiariotti2019acoustic}.

There are numerous algorithms that can be implemented to achieve beamforming. These algorithms can be categorised into time-domain beamforming such as Delay-Sum beamforming or frequency-domain beamforming such as Discrete-Fourier Transform (DFT) Single-Beam beamforming \cite{brandstein2013microphone,hamid2014performance}. Frequency-domain methods offer improved computational efficiency, signal fidelity, and robustness against two or more interfering sound sources in comparison to time-domain methods \cite{hamid2014performance,lockwood2004performance}. Though these are characteristics desirable for industrial applications, frequency-domain methods are less intuitive and more complex in their implementation. The processing of experimental data in this work has utilised time-domain methods as an initial method for validating the beamforming performance of the developed microphone array prototype. 

\begin{figure}[!ht]
    \centering
    \includegraphics[width=\linewidth,keepaspectratio]{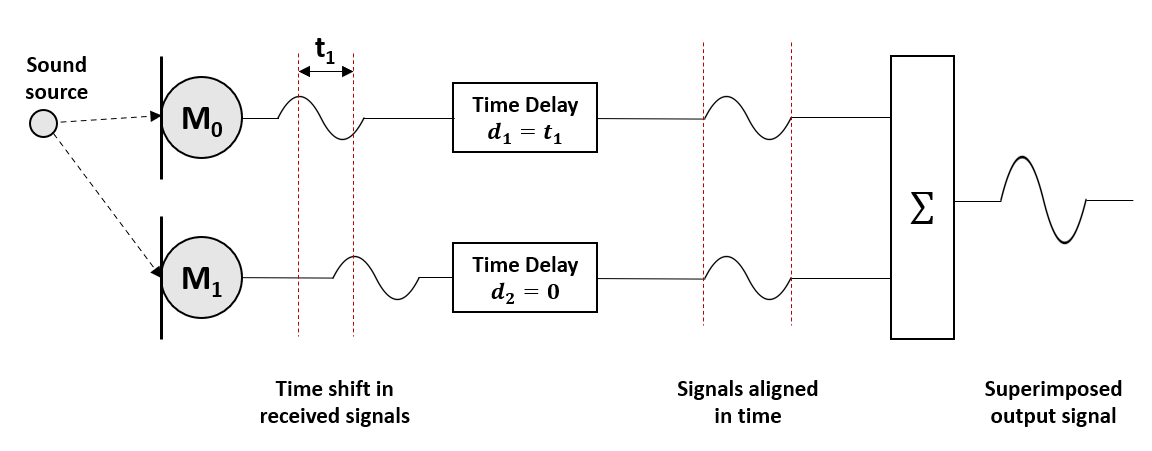}
    \caption{Block diagram of the time-domain delay-and-sum beamforming method which demonstrates time-shifting of sound data acquired by microphones in a ULA and its subsequent amplified output. (adapted from \cite{krol2015detecting})}
    \label{fig:Time-Domain DS Beamforming}
\end{figure}

The delay-and-sum (DS) beamformer is a standard time-domain technique that is widely documented in literature \cite{brandstein2013microphone,benesty2008microphone}. As illustrated in Figure \ref{fig:Time-Domain DS Beamforming} acquired sound signals are time-shifted for each sensor within a ULA based on the time difference of arrival between sensor $n$ and some reference sensor $M_2$. The summation of these time-shifted signals results in an effective amplification of the sound signal of interest and attenuation of unwanted noise through constructive and destructive interference respectively \cite{benesty2008microphone}. 

This forms the output of the DS beamformer from which the ULA's beam pattern can be generated. The beam pattern illustrates the performance of the ULA DS beamformer by visualising the estimated azimuth angle of a sound source relative to the position of the ULA. The angle difference between the estimated azimuth angle and the real azimuth angle provides a metric of performance measure by which the ULA can be assessed. 

A quantitative measure of beamforming performance for microphone arrays is its array gain. The Signal-to-Noise ratio (SNR) of a microphone measures the ratio between a reference signal level such as background noise to the noise level of the microphone output signal \cite{DigiKey2020MEMS}. By comparing the known SNR of a microphone in isolation to the output of the same microphone in an array, the array gain can be computed as follows \cite{brandstein2013microphone}:

\begin{equation}
    \label{Array Gain}
    \mathrm{Array \; Gain} = \frac{SNR_{Array}}{SNR_{Sensor}}
\end{equation}

Array gain quantifies the improvement in SNR between a reference sensor and the array output. Multiple publications quantify the beamforming performance of both digital and analog capacitive MEMS microphones using SNR gain. Examples include the investigation of the SNR gain of a 52-microphone array using digital output capacitive MEMS microphones \cite{tiete2014soundcompass} and a similar investigation using analog output capacitive MEMS microphones \cite{AnalogDevices}. However, there is a lack of published literature on the beamforming performance of digital output piezoelectric MEMS microphones. 
    
\subsection{Piezoelectric and Capacitive MEMS Microphones}
Micro-electromechanical Systems (MEMS) refer to microscopic devices comprised of moving mechanical elements and electrical components fabricated on a single silicon chip. MEMS sensors offer the advantage of smaller size, lower weight, less power demand, and greater versatility compared to their conventional sensor counterparts \cite{tanner2009mems}. In 2015 the global MEMS market was valued at USD 11.99 billion \cite{MEMSReport1} and is projected to achieve a CAGR of 6.34\% over the 2020 - 2025 period \cite{MEMSReport2}. This growth is driven by the increasing adoption of MEMS in consumer electronics devices, automotive and industrial applications, along with continual developments within the Internet of Things (IoT) space \cite{MEMSReport1}. 

\begin{figure}[!ht]
    \centering
    \includegraphics[width=\linewidth]{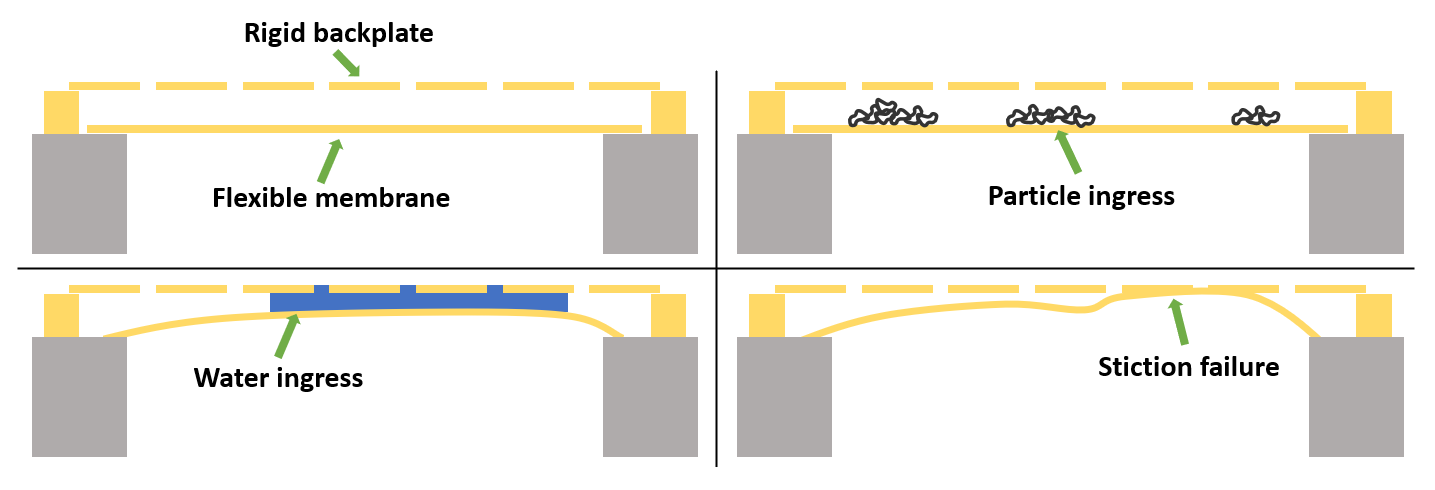}
    \caption[Failure modes of capacitive MEMS microphones - normal (upper left), particle ingress (upper  right), water ingress (lower left) and stiction failure (lower right)]{Failure modes of capacitive MEMS microphones: normal (upper left), particle ingress (upper  right), water ingress (lower left) and stiction failure (lower right). (adapted from \cite{Chew2017mems})}
    \label{fig:Capacitive MEMS Microphones Failure Modes}
\end{figure}

Capacitive MEMS microphones are the current standard of microphones used in beamforming. Their basic structure consists of an air-gap separated fixed backplate in parallel with a flexible membrane that vibrates in response to acoustic pressure. This produces a variation in the air gap, thereby altering the parallel plate capacitance and producing a voltage that corresponds to the incoming sound wave \cite{zawawi2020review}. The ability of the membrane to flex freely is critical to the performance of the device and can be compromised due to particle ingress, water ingress or mechanical shock  \cite{samper2003mems,Chew2017mems}. If the membrane is unable to flex freely, the microphone is no longer able to generate a capacitance proportional to incoming sound waves, thereby reducing performance or producing unusable data. The failure modes which can obstruct the motion of the membrane are illustrated in Figure \ref{fig:Capacitive MEMS Microphones Failure Modes} which compares the structure of a normal capacitive MEMS microphone against capacitive MEMS microphones under different failure modes. 

\begin{figure}[!ht]
    \centering
    \includegraphics[width=\linewidth]{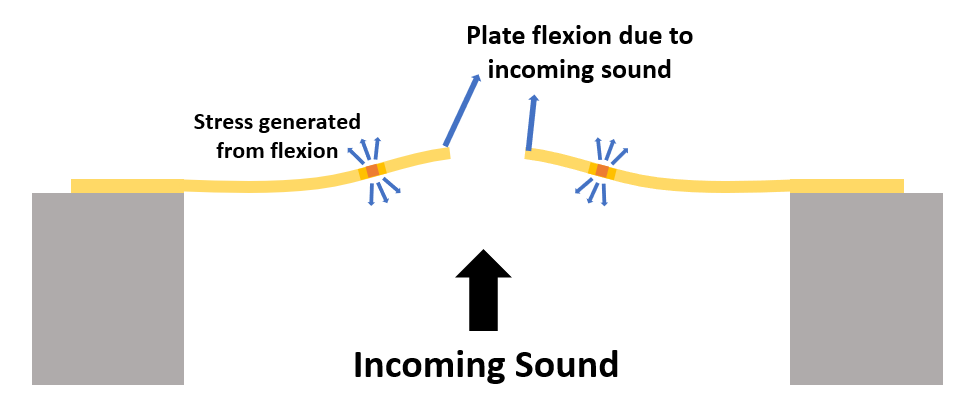}
    %\caption[Bottom view of the VM3000 piezoelectric MEMS microphone structure illustrating four cantilevered piezoelectric diaphragms (left), cantilever piezoelectric diaphragms flexing in response to incoming sound and generating stress within the plates which produces a corresponding electrical signal (right)]{Bottom view of the VM3000 piezoelectric MEMS microphone structure illustrating four cantilevered piezoelectric diaphragms (left), cantilever piezoelectric diaphragms flexing in response to incoming sound and generating stress within the plates which produces a corresponding electrical signal (right) \cite{Chew2017mems,Vesper2020VM3000}.}
    \caption[VM3000 cantilever piezoelectric diaphragms flexing in response to incoming sound and generating stress within the plates which produces a corresponding electrical signal (right)]{VM3000 cantilever piezoelectric diaphragms flexing in response to incoming sound and generating stress within the plates which produces a corresponding electrical signal (right). 
    (adapted from \cite{Chew2017mems,Vesper2020VM3000})}
    \label{fig:Vesper MEMS piezoelectric microphone}
\end{figure}

Piezoelectric MEMS microphones offer a potential solution to common failure modes of their capacitive counterpart. The basic structure of the VM3000 comprises of a cantilever diaphragm that flexes against acoustic pressure and generates a corresponding voltage through the piezoelectric effect \cite{grosh2013piezoelectric}. The piezoelectric cantilever plates remove the need for a backplate, thereby producing a design that is innately more robust against particle ingress, water ingress or mechanical shock. The lack of a backplate membrane ensures either particles or water are unable to remain embedded within the structure of the microphone, thereby ensuring performance is not inhibited even in the presence of particle or water ingress. Figure \ref{fig:Vesper MEMS piezoelectric microphone} illustrates the cantilever design of the VM3000 microphone. The microphone contains four cantilever structures that act as sound transducers against incoming sound. The stress generated by incoming sound is converted into a corresponding electrical signal.

Digital microphones, such as the VM3000, contain an integrated analog-to-digital converter (ADC) which enables digital signal processing to be conducted without the need for an external ADC. This allows the VM3000s to be interfaced directly with commercial off-the-shelf (COTS) hardware, such as the Raspberry Pi, making the construction of a small-scale self-contained microphone array capable of data acquisition, signal processing, and data transmission feasible.

%\subsection{Summary}
%Currently available literature does not explore the beamforming performance of digital piezoelectric MEMS microphones. Existing studies utilise capacitive MEMS microphone arrays for beamforming, which by design are less robust in comparison to piezoelectric MEMS microphones. This renders capacitive MEMS microphones unsuitable for application within harsh environmental conditions such as the resources industry. This work addresses this identified gap in literature by assessing the beamforming capabilities of the VM3000 digital piezoelectric MEMS microphones interfaced with COTS hardware.

\section{Hardware Selection and Design}
To test the beamforming capabilities of the VM3000 digital piezoelectric MEMS microphone, a microphone array prototype controlled through COTS hardware was developed. The design method is split into  hardware selection and design and software development.

The focus of hardware development is to produce a prototype that is capable of powering, interfacing with, and controlling the VM3000 microphones. 

\subsection{Digital MEMS Microphone}
The Vesper VM3000 microphone is an omnidirectional, bottom-port, Pulse Density Modulation (PDM) digital piezoelectric MEMS microphone. The PDM digital output enables the multiplexing of two microphones on a single data line. In an active state at 3.072MHz, the VM3000 consumes a maximum of 800$\mu$A and less than 1$\mu$A in sleep mode. It has a flat frequency response over 500 to 2000 Hz and an ingress protection rating of IP57, certified for dust and water ingress resistance \cite{Vesper2020VM3000}.

\subsection{Microcontroller}
The Raspberry Pi 4 Model B is chosen as the microphone array COTS controller. It is a 1.5GHz, quad-core CPU mini computer with wireless local area network (LAN) and Bluetooth 5.0 capabilities. It is capable of reading PDM digital data from two VM3000 MEMS microphones.  It is space-efficient with dimensions of 88mm x 58mm x 19.5mm and light-weight at 46g, characteristics which are desirable for interfacing with small-scale MEMS devices \cite{RPi42019datasheet}. Wi-Fi and Bluetooth capabilities enable the prototype to be accessed remotely and thereby operate as a stand-alone device.

\subsection{Hardware Design}
\begin{figure}[!ht]
    \centering
    \includegraphics[width=0.5\textwidth,keepaspectratio]{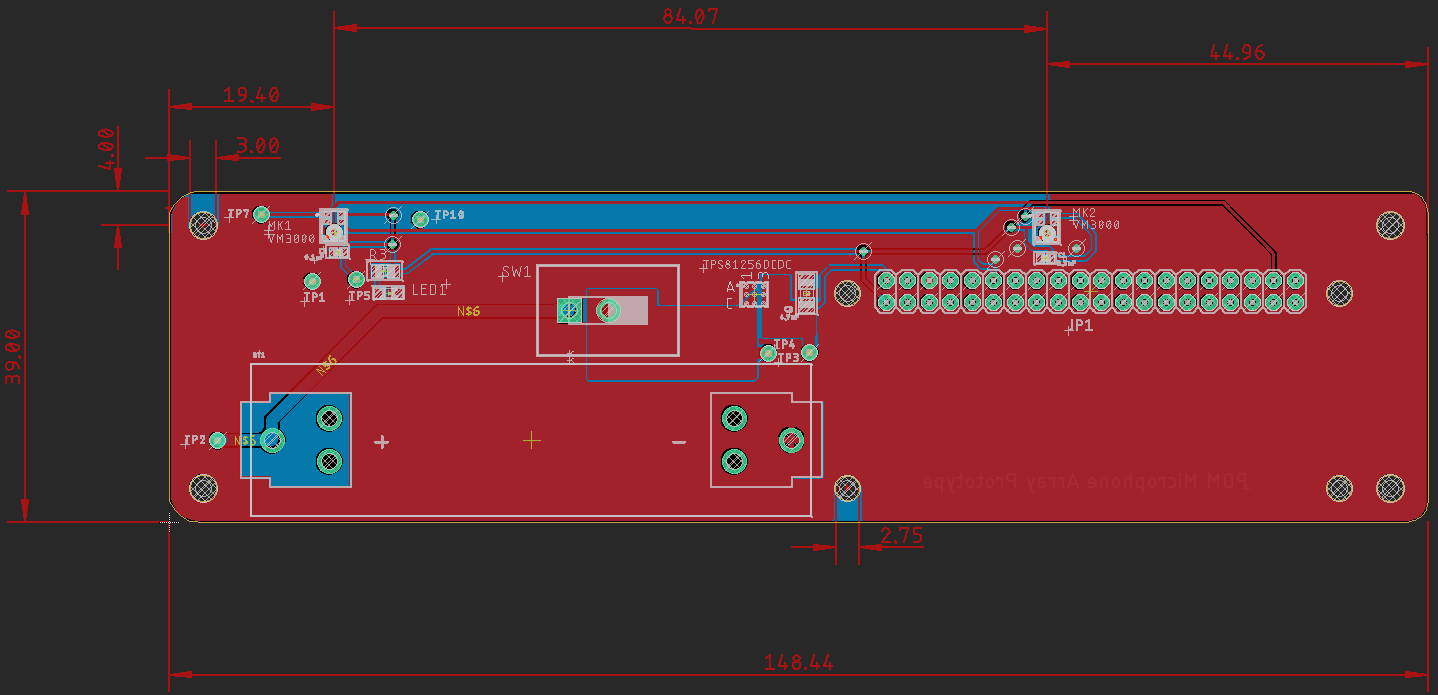}
    \caption{PDM microphone array prototype layout produced from Autodesk EAGLE. The array is designed for a LiPo 18650 battery cell, switch, TPS81256 boost converter, two VM3000 MEMS microphones and a 40-pin GPIO header.}
    \label{fig:PDM microphone array layout}
\end{figure}

\begin{figure}[!ht]
    \centering
    \includegraphics[angle=90,width=0.5\textwidth,keepaspectratio]{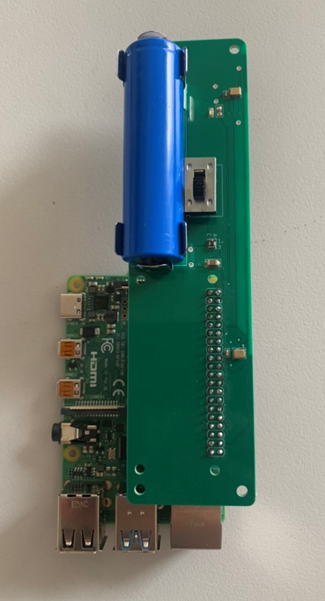}
    \caption{Top view of a completed, fully populated 2-element linear microphone array prototype connected to a Raspberry Pi 4 Model B. (NB: The Pi 4 Model B is unable to be powered from a single 18650 battery)}
    \label{fig:2-element linear microphone array}
\end{figure}

Design of the microphone array prototype PCB is performed using EAGLE and the completed board is illustrated in Figure \ref{fig:PDM microphone array layout} and Figure \ref{fig:2-element linear microphone array}. 
The PCB design contains a LiPo 18650 battery cell and TPS81256 boost converter.
These components where to be utilised when the array was controlled by a Raspberry Pi Zero, but became unnecessary once the Pi Zero was replaced with a Pi 4 Model B, as discussed in Section~\ref{PiProblems}.
The primary objective of this design is to produce a microphone array prototype of minimal size to ensure manufacturing costs remain low and portability is maintained. Inter-microphone spacing is the key measure that determines the physical size of the microphone array prototype PCB. A maximum inter-microphone spacing of 8.4 cm is chosen through a compromise of the physical PCB dimensions, tolerable output signal pitch during experimental testing, and the flattest frequency response range of the VM3000 microphones (500 to 2000 Hz). 
By taking the upper bound of this frequency range and using equation \ref{Inter-microphone distance restriction}, the maximum inter-microphone spacing is determined to be approximately 8.6 cm.
Therefore the designed inter-microphone spacing of 8.4 cm falls within the acceptable spacing distance, ensuring no spatial aliasing occurs when recording signals within 500 to 2000 Hz. 

\section{Software Design}
The focus of software design is to develop a remote process for data acquisition, data transfer, and signal processing using a laptop or equivalent device. This involves the simultaneous control of lab equipment, to generate test signals, and the Raspberry Pi on the microphone array prototype, for sound recordings. A high-level overview of the software operation is as follows:

\begin{enumerate}
    \item Import modules to be used in the script
    \item Establish connection with a Keysight 33500B Waveform Generator, either via wireless local area network if the signal generator is Ethernet-enabled or via direct USB connection with a laptop.
    \item Establish connection with Raspberry Pi on microphone array prototype via SSH protocol.
    \item Load class of test functions which will control the test signal output, sound recording duration of the microphone array prototype, and data transfer of .WAV recordings to laptop or equivalent device.
    \item Perform time-domain delay-and-sum beamforming using the .WAV recordings and using theoretical data.
    \item Visualise beam pattern of real data and theoretical data on the same plot.
\end{enumerate}

\section{Microphone Array Prototype Build}
\subsection{Hardware Implementation}
\subsubsection{Hardware Assembly}
A combination of reflow soldering for SMD components and hand iron soldering for through-hole components is used to assemble the microphone array prototype. Due to the small footprint of the VM3000 MEMS microphones, TPS81256 boost converter, and other SMD components, a solder paste stencil are used to lay solder paste accurately onto the pads of SMD components. These components are mounted via reflow soldering using a hand-held heat gun. Through-hole components such as the LiPo cell battery clips, switch and female 40-pin general-purpose input/output (GPIO) header are assembled by hand using a soldering iron. Since the female 40-pin GPIO sits on the underside of the microphone array prototype, any Raspberry Pi with a male GPIO header will be able to interface directly with the prototype. 

\subsubsection{Raspberry Pi Zero W and Raspberry Pi 4 Model B performance}\label{PiProblems}
\begin{figure}[!ht]
    \centering
    \includegraphics[width=\linewidth,keepaspectratio]{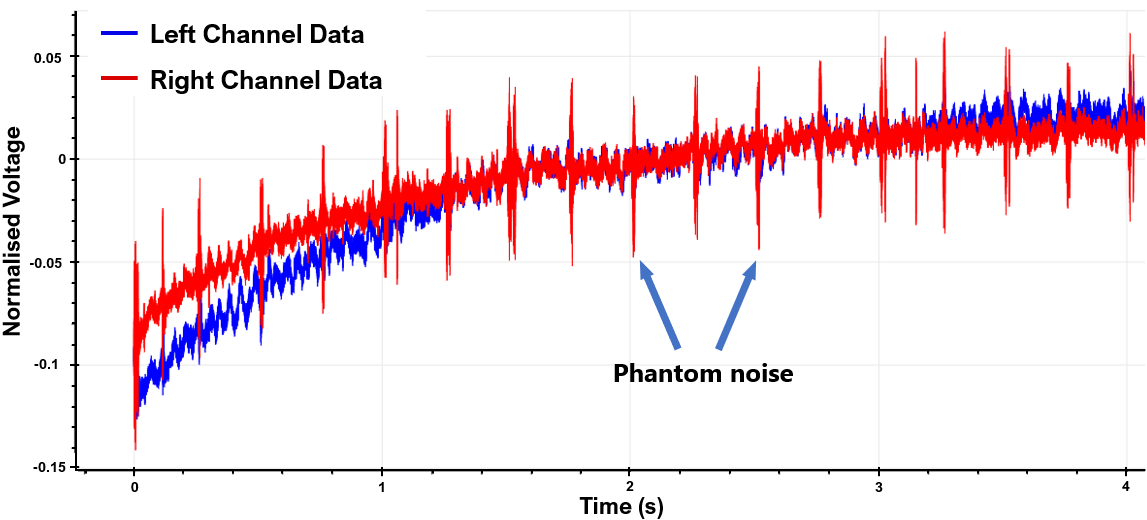}
    \caption{Phantom noise observable in the raw .WAV file through distinct, periodic peaks during ambient noise recording when the microphone array prototype is interfaced with the Raspberry Pi Zero W.}
    \label{fig:Phantom noise}
\end{figure}

Initial development was done using a Raspberry Pi Zero W. Interfacing the microphone array prototype with the Raspberry Pi Zero W produced phantom noise in sound recordings. Two separate builds of the microphone array prototype were tested for different input signals. The phantom noise is present when the microphone array is interfaced with the Raspberry Pi Zero W and powered using either the 18650 battery cell or the micro USB power input. Sampling frequency was adjusted to no effect. In all cases, a rhythmic, audible ticking noise in both left and right channel data is audible when sound recordings are replayed. Figure \ref{fig:Phantom noise} illustrates this observable phantom noise through a plot of a .WAV file from an ambient noise recording. The phantom noise is characterised by periodic peaks within the .WAV file which is observable in both left and right channel data. These peaks occur at the same points in time and at the same frequency in both left and right channel data.  

\begin{figure}[!ht]
    \centering
    \includegraphics[width=\linewidth,keepaspectratio]{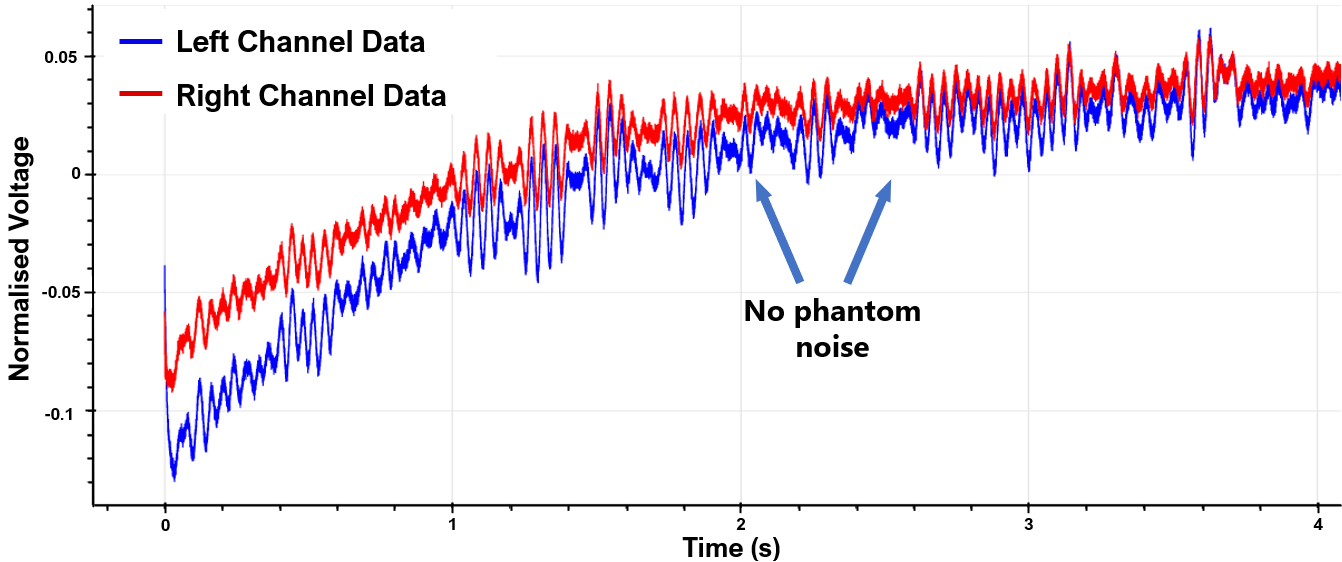}
    \caption{Phantom noise no longer observable in the raw .WAV file during ambient noise recording as periodic peaks are no longer present when the microphone array prototype is interfaced with the Raspberry Pi 4 Model B.}
    \label{fig:No Phantom Noise RPi 4}
\end{figure}

This phantom noise is no longer observable when the microphone array prototype is interfaced with the Raspberry Pi 4 Model B with an official power supply. Figure \ref{fig:No Phantom Noise RPi 4} illustrates an ambient noise recording when the microphone array prototype is interfaced with a Raspberry Pi 4 Model B. The distinct periodic peaks of the phantom noise are no longer observable within the .WAV file nor is it audible when listening to replays of sound recordings. 

It is likely that the phantom noise originates internally to the Raspberry Pi Zero W. A possible cause is a lack of hardware processing power in the Raspberry Pi Zero W for high frequency data sampling at 96 kHz. The  Raspberry Pi 4 Model B contains a quad-core CPU, 1.5GHz clock, and 8GB RAM whilst the Raspberry Pi Zero W contains a single-core CPU, 1GHz clock, and 512MB RAM. Since the Raspberry Pi Zero W is a single-core CPU device, it must share processing power for PDM data with its other running processes. This inability to multitask may result in some interaction between PDM recording and background processes which corrupts the .WAV file data. Further work is required to investigate this observation.

%\subsection{Power Source}
%The Polymer Lithium-Ion (LiPo) 18650 battery cell (3.7V 2600mAh) is used as the wireless, rechargeable power source for the microphone array. A reliable, rechargeable power source is essential to achieving a stand-alone prototype device that is mobile and free of external cabling. The chosen 18650 battery cell is simple to use and is capable of handling approximately 300 charge cycles. A single LiPo cell is sufficient to fully power the Raspberry Pi Zero W for an extended duration. However, a single LiPo cell is not capable of powering the wireless LAN of the Raspberry Pi 4 Model B. Therefore, 

%\subsection{Supporting Hardware}
%The TPS81256 boost converter is chosen as the DC-to-DC voltage step-up converter. It is used to step up the 3.7V LiPo cell voltage to the required 5V input voltage for the Raspberry Pi controller. A single-pole single-throw switch sits in between the LiPo cell and the TPS81256 boost converter to provide a manual method for controlling power to the controller. An LED is installed to provide a visual indicator for power flow through the controller and microphones. Where necessary, Surface Mount Device (SMD) capacitors are implemented to provide output capacitance to reduce transient voltage fluctuations.

\subsection{Software Implementation}
\subsubsection{Software Development}
A central code base capable of controlling the Raspberry Pi 4 Model B, commanding laboratory test equipment, transporting data remotely, and conducting signal processing is developed specifically for this project by Ricky Leman \cite{UWASHLGitHubVM3000}. 
The Python implementation of the time-domain delay-sum beamforming and visualisation of experimental data is developed by Dr. Ben Travaglione. 
It is a simplistic implementation of the time-domain delay-sum beamforming algorithm based on fundamental theoretical principles. It is capable only of modelling the theoretical beam pattern of a single, perfect sound source and a single receiver in infinite space. 
Jupyter Lab is used as an open-source web-based tool for developing the project code. 
Python is chosen as the programming language of choice due to its ease of implementation and access to packages relevant to this work. 
The source code for this work is available on the UWA System Health Lab GitHub \cite{UWASHLGitHubbeamforming}. 

Key Python packages that have been used include:
\begin{enumerate}
    \item vxi11 - This package supports the VXI-11 Ethernet instrument control protocol for controlling VXI11 and LXI compatible instruments. It enables Ethernet control of the Keysight 33500B Waveform Generator.
    \item pyvisa - This package enables control of measurement devices independently of its interface. It enables USB control of the Keysight 33500B Waveform Generator.
    \item paramiko - Paramiko is a Python implementation of the SSHv2 protocol which  provides client and server functionality. It enables remote SSH access to the Raspberry Pi 4 Model B.
    \item scp - This package is used in conjunction with paramiko to send and receive files via the scp1 protocol. It enables transfer of sound files locally stored on the Raspberry Pi 4 Model B to the user's device.
    \item bokeh - Bokeh enables the creation of interactive data visualisation on web browsers. It enables raw .WAV file data to be visualised in an interactive plot which dynamically updates based on changing inputs or user interaction{.}
    
\end{enumerate}

\subsubsection{Enabling PDM audio on Raspberry Pi Broadcom System on Chip}
Interfacing PDM audio with the Raspberry Pi range involves altering the Broadcom System on Chip (SoC) source code. 
The BCM2835(2012) and BCM2711(2020) ARM Peripherals datasheets provide technical information regarding the PDM input mode operation but do not provide detailed instructions on how to activate the PDM input mode. 
In December 2020, an online review had yielded a single set of high-level instructions on the Raspberry Pi forums and no formal documented process. 
A reproducible method has been developed for this paper and documented on The UWA System Health Lab GitHub site\cite{UWASHLGitHubVM3000}, as well being as shared on public Raspberry Pi forums. 
This method has been tested and verified to be successful on the Raspberry Pi 4 Model B.
The method also enables PDM audio on the Raspberry Pi Zero W, however there are unresolved recording issues, as discussed in Section~\ref{PiProblems}.

\section{Experimental Method}
%\subsection{Testing Goals}
The goal of this testing is to compare the accuracy of real beam patterns created by sound data acquired using the VM3000 MEMS microphones against theoretical beam patterns generated using idealised data. Differences between peak angles and overall shape are the characteristics of interest as they determine the beamforming accuracy of the microphone array prototype and demonstrate the effects of sound interference.

\subsection{Experiment Setup}
\begin{figure}[!ht]
    \centering
    \includegraphics[width=\linewidth,keepaspectratio]{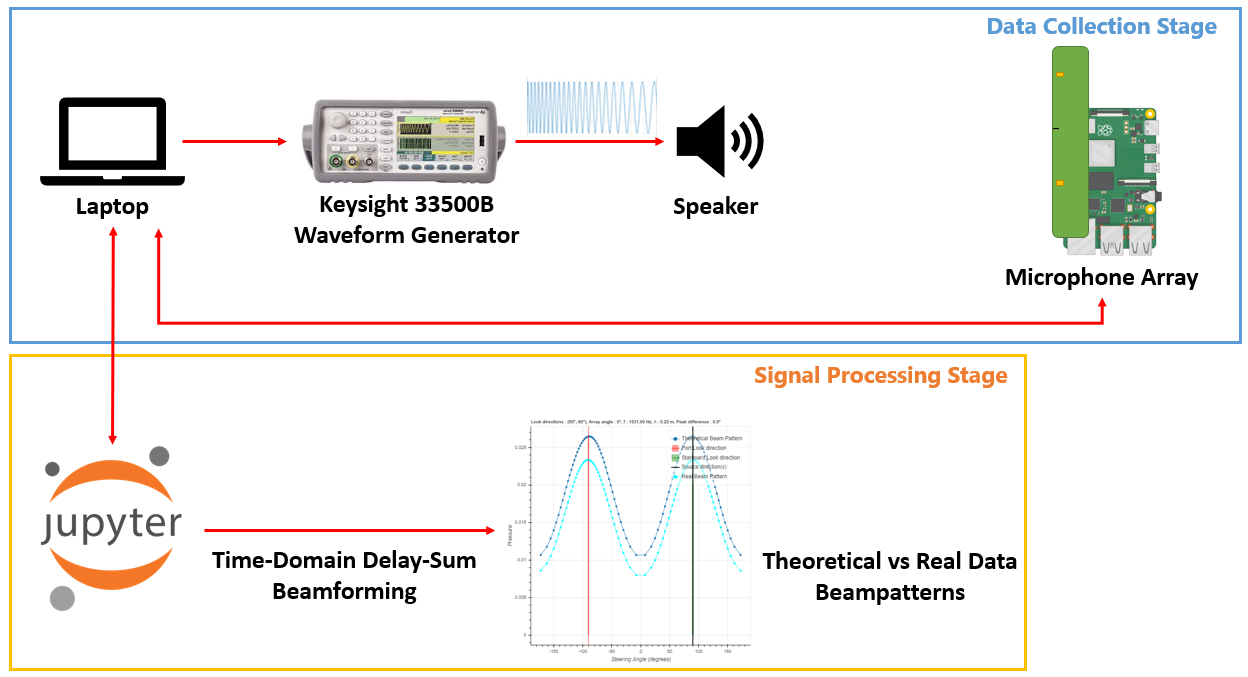}
    \caption{Two-stage experiment setup for beamforming testing of the microphone array prototype. The microphone array records a 500 - 3000 Hz linear frequency sweep generated by a Keysight 33500B Waveform Generator fed into a speaker. The microphone array is rotated by $20\degree$ increments between $0\degree$ and $360\degree$. Time-Domain Delay-Sum beamforming is conducted on collected data through Python to generate theoretical and real beampatterns.}
    \label{fig:Experiment setup for beamforming testing of VM3000 microphones}
\end{figure}

A repeatable testing sequence is developed to ensure consistency in data collection. The experiment setup consists of two distinct stages which are illustrated in Figure \ref{fig:Experiment setup for beamforming testing of VM3000 microphones}. Testing is conducted in the UWA Anechoic chamber.

\begin{figure}[!ht]
    \centering
    \includegraphics[angle=270,width=0.5\textwidth,keepaspectratio]{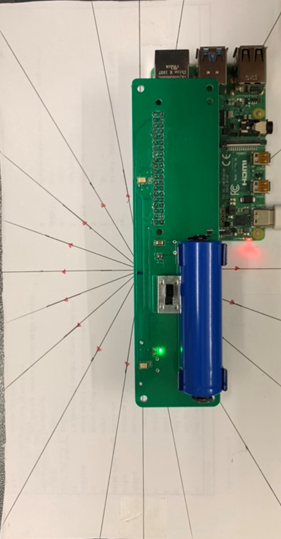}
    \caption{Angle bearing guide with increments of $20\degree$ used for rotating the microphone array prototype about its inter-microphone centre for each iteration of testing.}
    \label{fig:Anglebearing}
\end{figure}

%\subsubsection{Data Collection Stage}
The microphone array prototype is placed at a starting position perpendicular to the speaker (array angle of $0\degree$) along the same horizontal plane. 
A Python script running in a Jupyter Lab notebook is used to command a Keysight 33500B Waveform Generator to output a linear frequency sweep between 500 Hz to 3000 Hz through the speaker. 
This frequency range is chosen to observe the beamforming performance for the operating design range of 500 to 2000 Hz and to observe the effects of aliasing for frequencies greater than 2000 Hz. 
The output signal amplitude is kept constant for all tests. 
The linear sweep output is repeated for each $20\degree$ array angle rotation until 18 readings corresponding to a $360\degree$ rotation is recorded. The angle bearing guide used for this is illustrated in Figure~\ref{fig:Anglebearing} and was constructed manually using a protractor. 
Experimental data is collected in .WAV format using the Linux \texttt{arecord} function via the Raspberry Pi 4 Model B.

\subsubsection{Signal Processing Stage}
Python running on Jupyter Lab is used for signal processing and data visualisation. Both experimental and theoretical data is processed using time-domain delay-sum beamforming. Theoretical beam patterns are generated assuming a completely anechoic environment free of reflections and a single, perfect incoming sound signal. Real beam patterns are generated using experimental data.

For assessing a single combination of array angle and frequency, a normalised theoretical and real beam pattern is generated and overlayed on a single plot. This allows for the comparison of peak angle and shape differences. 

For assessing all combinations of array angles and frequencies, the area difference between normalised theoretical and real beam patterns is computed and plotted on a heatmap. The accuracy of the microphone array prototype beamforming performance is quantified by calculating the Root Mean Square Error (RMSE) between the real and theoretical beam patterns generated by the dataset.

\subsubsection{Anechoic environment setup}
The test is conducted in the UWA Anechoic Chamber. This environment is chosen to evaluate the microphone array prototype beamforming performance under anechoic conditions, enabling like-for-like comparison with the theoretical beam pattern data. The test environment is lined with sound-absorbing materials such as curtains for the walls and foam padding for the floors underneath the metal grating. These materials prevent sound reflections from occurring. At the time of testing, a wooden wall had been installed for other works within the anechoic chamber, introducing a surface for sound reflections. The impact of this is minimised by placing the microphone array prototype in a corner surrounded by sound-dampening surfaces.

\section{Results and Discussion}
\begin{figure}[!ht]
    \centering
    \includegraphics[width=100mm,keepaspectratio]{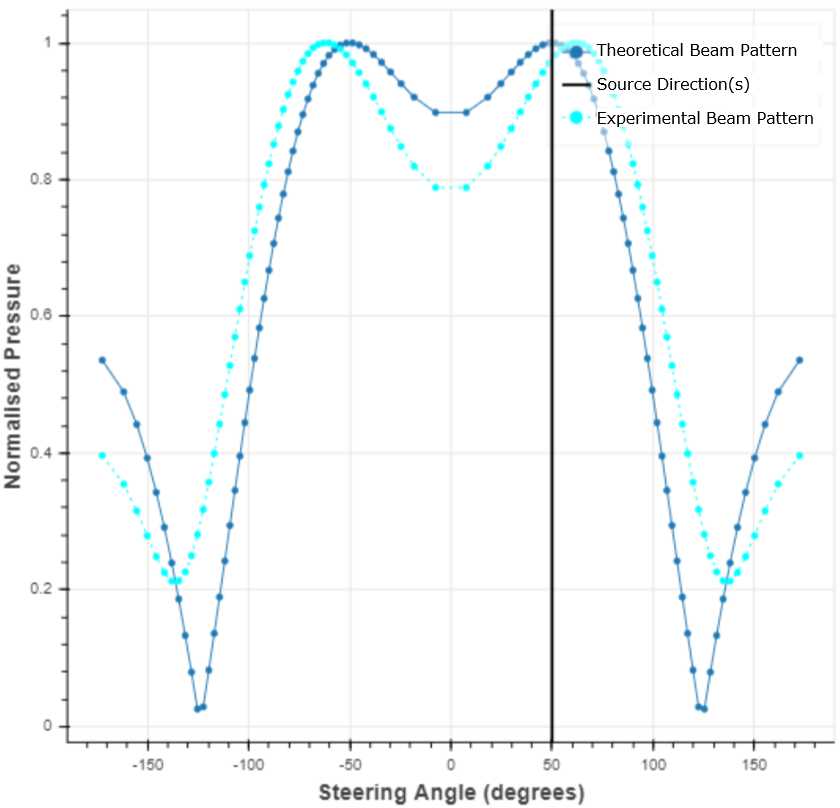}
    \caption{Accurate beamforming performance as seen by a peak angle difference of approximately $10\degree$ for data collected by the microphone array prototype in an anechoic test environment at an array angle of $50\degree$ and frequency of 1650 Hz.}
    \label{fig:beam pattern-anechoic}
\end{figure}

%\subsection{Test: Anechoic environment}
Three trials of experimental data were gathered under the same test conditions to minimise the impact of random errors in the anechoic chamber. To ensure the effects of sound reflections remain consistent between trials, all test apparatus such as the ladder and crates were kept in place between trials. Only the microphone array prototype itself is rotated between each recording. Sound data acquired by the microphone array prototype in an anechoic environment demonstrates a beam pattern which agrees with the developed theoretical model. Figure~\ref{fig:beam pattern-anechoic} illustrates the microphone array prototype's beam pattern at an array angle of $50\degree$ and frequency of 1650 Hz generated using data collected in a single trial. The solid, blue line represents the theoretical beam pattern generated using idealised sound data. The dotted, cyan line represents the experimental beam pattern generated using collected test data by the microphone array and the solid, black line represents the source angle direction. No aliasing is present within the experimental beam pattern, however, an approximately $10\degree$ mismatch in peak angles is observable. This peak angle mismatch is likely to be resultant of surface reflections by the wooden panel installed within the anechoic chamber. Reflections from the microphone array hardware itself and the wooden panel upon which it is resting may also be contributing factors.

\begin{figure}[!ht]
    \centering
    \includegraphics[height=6cm,width=100mm]{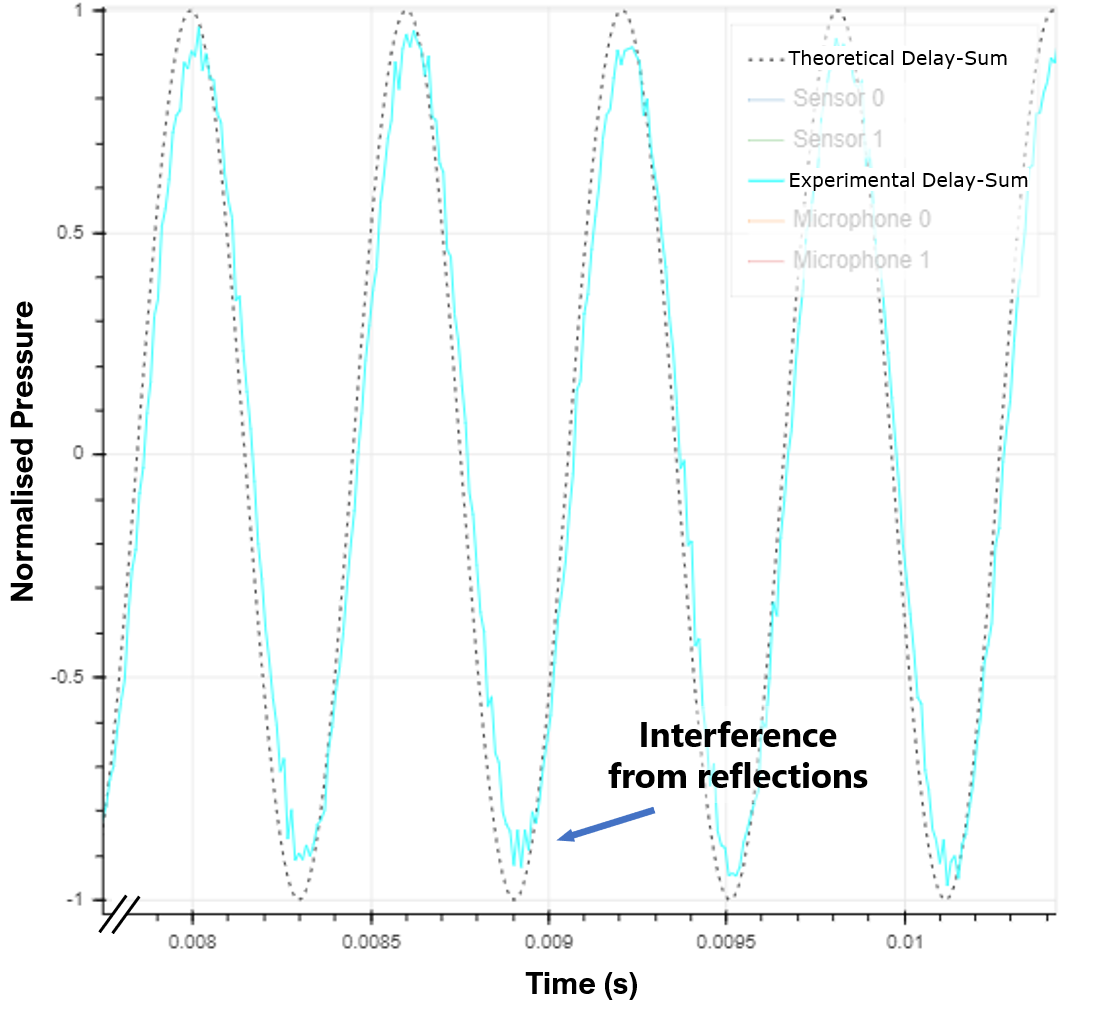}
    \caption{Sample of time-domain delay-sum beamforming (cyan) compared with the theoretical delay-sum beamforming  (black dots).
    Differences in these signals is likely due surface reflections in the UWA Anechoic Chamber.}
    \label{fig:noiseinanechoicchamber}
\end{figure}

Figure \ref{fig:noiseinanechoicchamber} compares the theoretical delay-sum waveform with the experimental delay-sum waveform. 
Sound reflections and noise are observable within the experimental delay-sum waveform signal characterised by perturbations within the sinusoid. 
This suggests an imperfect anechoic environment as discussed and is likely to be a large contributing factor to peak angle mismatches in conjunction with the array's physical structure. 

The theoretical and experimental beam patterns change as a function of array angle and frequency. 
As we are interested in both the peak angle difference and shape difference between theoretical and experimental beam patterns, finding the normalised area difference between the two beam patterns captures both performance measures in a single value. 
A small normalised area difference indicates the experimental beam pattern matches the shape of the theoretical beam pattern well, therefore suggesting strong beamforming performance. 
A large normalised area difference indicates the shape of the experimental beam pattern is a mismatch with the theoretical beam pattern, thereby suggesting poor beamforming performance.

\begin{figure}[!ht]
    \centering
    \includegraphics[width=100mm,keepaspectratio]{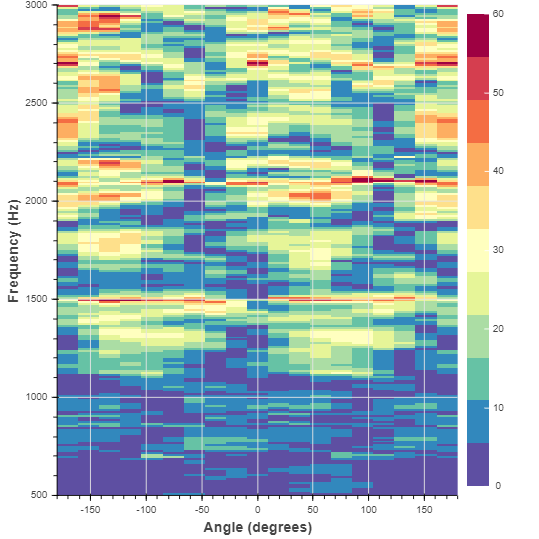}
    \caption{Normalised area difference between theoretical and real beam patterns for all array angle and frequency combinations for test data collected in an anechoic environment. 
    The heatmap shows agreement between experimental and theoretical beam patterns for the majority of array angle and frequency combinations, indicating accurate beamforming performance.}
    \label{fig:Heatmapanechoic}
\end{figure}

Figure~\ref{fig:Heatmapanechoic} visualises the normalised area difference between theoretical and experimental beam patterns for all array angles and frequencies recorded. 
This enables the beamforming performance of the microphone array prototype across all combinations of array angles and frequencies to be observed. 
These values have been mapped to a colour scale from purple to red. 
Purple represents the smallest area difference and therefore the best beamforming performance. 
Whereas red represents the largest area difference and hence the worst beamforming performance. 

A strong match in shape between experimental and theoretical beam patterns is observable, indicative of strong beamforming performance across all combinations of array angles and frequencies tested for an anechoic environment. 
The beamforming performance of the microphone array prototype is quantified by computing the average RMSE between experimental and theoretical beam patterns for each data set generated from each trial between the frequency operating range of 500 - 2000 Hz. 
The average RMSE values for each trial, and the overall average are shown in Table~\ref{RMSEvalues}.

\begin{table}[ht]\centering
\begin{tabular}{l|l|l|l|l|}
\cline{2-5}
                                         & Trial 1 & Trial 2 & Trial 3 & Overall \\ \hline
\multicolumn{1}{|l|}{Average RMSE value} & 12.52\% & 10.33\% & 10.45\% & 11.10\% \\ \hline
\end{tabular}
\caption{Average RMSE values from each trial between the frequency operating range of 500 - 2000 Hz.}\label{RMSEvalues}
\end{table}

The experimental beampattern generated by data collected from the microphone array prototype produces an RMSE of 11.10 \% when compared to theoretical beam patterns. 

When the signal of interest exceeds 2000 Hz, a decrease in beamforming performance is observable. 
This observation agrees with theory as spatial aliasing is expected to occur for signal frequencies that exceed the microphone array prototype design frequency limit of 2000 Hz. 
However, for signals below 2000 Hz, the heatmap indicates the majority of array angles and frequencies produce experimental beam patterns which accurately match with theoretical beam patterns. 
Hot spots in Figure \ref{fig:Heatmapanechoic} below 2000 Hz indicate a deviation between theoretical and experimental beam patterns. 
These are likely the result of sound reflections from the wooden wall interfering with the signal of interest and also due to obstruction of sound by the asymmetric microphone array prototype structure. 
Therefore, the developed microphone array prototype exhibits a strong beamforming performance in anechoic environments. 

\section{Conclusion}
This paper demonstrates the design, build and test of a stand-alone digital two-element microphone array prototype using the VM3000 MEMS microphone. 
The prototype is tested in an anechoic environment. 
Time-domain delay-and-sum beamforming is implemented using Python in a Jupyter notebook. 
Test results in anechoic environments demonstrate that the microphone array prototype exhibits accurate beamforming when subject to an environment with minimal surface reflections. 
This suggest the VM3000 MEMS microphones are capable of beamforming accurately under conditions of minimal sound reflections, however further work is required to develop a microphone array using the VM3000 MEMS microphones that is suitable for real-world applications.

%%%%%%%%%%%%%%%%%%%%%%%%%%%%%%%%%%%%%%%%%%
%\authorcontributions{
%For research articles with several authors, a short paragraph specifying their individual contributions must be provided. The following statements should be used ``
%Conceptualization, Ben Travaglione; methodology, Ricky Leman, Ben Tavaglione; software, Ricky Leman, Ben Tavaglione; validation, Ricky Leman, Ben Tavaglione; formal analysis, Ricky Leman, Ben Tavaglione.; %investigation, Ricky Leman, Ben Tavaglione.; 
%resources, Melinda Hodkiewicz; data curation, Ricky Leman, Ben Tavaglione; writing---original draft preparation, Ricky Leman; writing---review and editing, Ben Travaglione, Melinda Hodkiewicz. 
%;  supervision, X.X.; project administration, X.X.; funding acquisition, Y.Y. 
%All authors have read and agreed to the published version of the manuscript.}
%'', please turn to the  \href{http://img.mdpi.org/data/contributor-role-instruction.pdf}{CRediT taxonomy} for the term explanation. Authorship must be limited to those who have contributed substantially to the work~reported.}

%\funding{Please add: ``This research received no external funding'' or ``This research was funded by NAME OF FUNDER grant number XXX.'' and  and ``The APC was funded by XXX''. Check carefully that the details given are accurate and use the standard spelling of funding agency names at \url{https://search.crossref.org/funding}, any errors may affect your future funding.}

%\institutionalreview{Not applicable.}

%\informedconsent{Not applicable.}

The data presented in this study are available at \cite{UWASHLGitHubVM3000}.

\section*{Acknowledgments}
Melinda Hodkiewicz acknowledges support from the BHP Fellowship for Engineering for Remote Operations which supports the UWA System Health Lab work.

The authors declare no conflict of interest.

%Bibliography
\bibliographystyle{unsrt}  
\bibliography{references}  

%%%%%%%%%%%%%%%%%%%%%%%%%%%%%%%%%%%%%%%%%%
%% Optional
%\appendixtitles{no} % Leave argument "no" if all appendix headings stay EMPTY (then no dot is printed after "Appendix A"). If the appendix sections contain a heading then change the argument to "yes".
%\appendixstart
%\appendix


%\reftitle{References}
%\externalbibliography{yes}
%\bibliography{references}
%%%%%%%%%%%%%%%%%%%%%%%%%%%%%%%%%%%%%%%%%%
%% for journal Sci
%\reviewreports{\\
%Reviewer 1 comments and authors’ response\\
%Reviewer 2 comments and authors’ response\\
%Reviewer 3 comments and authors’ response
%}
%%%%%%%%%%%%%%%%%%%%%%%%%%%%%%%%%%%%%%%%%%
\end{document}